\documentclass[pra,bibnotes]{revtex4}
\usepackage{graphicx}
\begin{document}
\draft
\def\ds{\displaystyle}
\title{ Anomalous features of non-Hermitian topological states }
\author{C. Yuce }
\address{Department of Physics, Eskisehir Technical University, Eskisehir, Turkey }
\email{cyuce@eskisehir.edu.tr}
\date{\today}
\begin{abstract}
Topological states in non-Hermitian systems are known to exhibit some anomalous features. Here, we find two new anomalous features of non-Hermitian topological states. We consider a one dimensional nonreciprocal Hamiltonian and show that topological robustness can be practically lost for a linear combination of topological eigenstates in non-Hermitian systems due to the non-Hermitian skin effect. We consider a two dimensional non-Hermitian Chern insulator and show that chirality of topological states can be broken at some parameters of the Hamiltonan. This implies that the topological states are no longer immune to backscattering in 2D.
\end{abstract}
\maketitle

\section{Introduction}

Topological band theory and its  non-Hermitian extension have attracted great attention in the past decades \cite{sondeney1,aah1,ghatakdas,1d5,ann01,eklon}. The periodic table of Hermitian topological insulators is well-known, but its non-Hermitian counterpart has not yet been fully constructed. Fortunately, some recent attempts pave the way for constructing it \cite{ptTI,cyek2}. Non-Hermitian topological edge states were initially found in simple models such as complex extensions of the Su-Schrieffer-Heeger (SSH) model \cite{1d1,1d2,1d6,1d7,1d8,1d9,1d10,1d11,1d3ekl,1d12,1d13,1d14,1d3,1d15,1d16,floquet1,floquet2,bhjkl,feng,ann2} and Kitaev model \cite{kita1,kita2,kita3,kita4,kita5}. Then more complex models has been introduced and studied in the literature \cite{cyek1,cyek3,cyek4,cyek5,cyek6,genel000,genel001,genel002,genel003,genel004,genel005,genel006,genel007,genel008,genel009,genel0010,genel0011,genel0012,genel0013,genel0014,genel0015,genel0016,genel0017,genel0019,genel0020,genel0021,takata,ann3,ann4,ann5,ann6,ann7}. In a recent paper, it was shown that topological phase can also arise in a non-Hermitian quasicrystal with parity-time symmetry \cite{nhquasi}. Note that topological edge states in a non-Hermitian system can have real or complex energy eigenvalues. The latter one can be used as a topological laser \cite{laser} or spontaneous topological pump at large times \cite{yucepump}. \\
One of the main problem in the theory of non-Hermitian topological systems is to understand the bulk-boundary correspondence \cite{bulkboun01,bulkboun13,ueda1a,bulkboun02,bulkboun02b,bulkboun04b,bulkboun06,bulkboun07,bulkboun08,bulkboun09,bulkboun10,bulkboun11,bulkboun12,bulkboun14,bulkboun15,bulkboun16}. Unfortunately, the standard bulk-boundary correspondence fails in many non-Hermitian systems. As opposed to Hermitian systems, topological phase transition points can not be generally determined using periodical form of the insulating non-Hermitian Hamiltonian. This is because of the fact that energy eigenvalues of topological and bulk states can depend sensitively on boundary conditions. Therefore, one must study non-Hermitian systems with open edges to precisely explore topological states. Furthermore, the non-Hermitian topological invariants explored so far in the literature are generally model dependent in sharp contrast to Hermitian systems \cite{winding1,winding3,yenice}. In other words, one can find at least two different topological numbers in the literature that can predict different topological phase diagram for a given non-Hermitian gapped Hamiltonian. In \cite{pseudo}, the existence of topological edge states for the SSH model with gain and loss was explained using the idea of pseudo topological insulator. The so called non-Hermitian skin effect was introduced in a nonreciprocal tight binding lattice with asymmetrical couplings \cite{bulkboun03,bulkboun04}. It states that not only topological states but also bulk states are localized around either edge, which causes the density of states at the edge to be increased. The accumulation of the eigenstates around one edge  can be understood as an amplification of them in one way and a corresponding decaying in the opposite way due to an imaginary gauge field \cite{bulkboun07}. Recently, the existence of hybrid skin-topological modes in a 2-dimensional system \cite{skin01} and non-Hermitian anomalous skin effect have been predicted \cite{anaskef,robustbulks}. In the latter case, topological states becomes extended all over the system. This has no analog in Hermitian systems as Hermitian topological states always occur around the edges where topological phase transition occurs.\\
It is well known that topological states in non-Hermitian systems show some anomalous features \cite{bulkboun01,bulkboun13,ueda1a}. In this paper, we explore some other anomalous features of non-Hermitian topological states. These are the superposition-induced loss of topological protection and chirality breakdown in a two dimensional non-Hermitian Chern insulator. Any linear combination of topological states is certainly topological in Hermitian systems. Here we find a one dimensional nonreciprocal Hamiltonian and show that a symmetric combination of its topological edge states is not topologically protected and grows unboundedly in time. As another anomalous feature, we consider a two dimensional non-Hermitian system. In a two dimensional Hermitian Chern insulator, topological edge states are protected and hence backscattering of topological edge states from symmetry protecting perturbative disorders are forbidden. We show that this is not always the case in non-Hermitian systems since the chirality can be broken due to the non-Hermitian skin effect.

\section{Fragile topological zero energy state}

Consider the 1D SSH tight-binding chain with nonreciprocal hopping amplitudes for the first sub-lattice. The non-Hermitian Hamiltonian under the periodic boundary condition can be written as
\begin{eqnarray}\label{kiRYazm}
\mathcal{H} (k)=\left(\begin{array}{cc} 0& t_1+{t_2}~e^{-ik} \\    t_1^{\prime}+t_2~e^{ik}  & 0 \end{array}\right)
\end{eqnarray}
where the real-valued parameters $t_1$, $t_2$ and $t_1^{\prime}$ are  hopping amplitudes. The Hamiltonian is Hermitian if $t_1={t_1}^{\prime}$. One can easily see that this Hamiltonian has chiral symmetry: $\ds{\sigma_z\mathcal{H} (k) \sigma_z=-\mathcal{H} (k) }$ where $\sigma_i$ refers to Pauli matrices. This implies that eigenvalues come in pairs at a given $k$. They are given by $E_{\mp}=\mp\sqrt{( t_1+{t_2}~e^{-ik})  (   t_1^{\prime}+t_2~e^{ik}  )   }$. The corresponding non-Hermitian winding number can be found in \cite{genel0019,winding1}. We refer the reader to \cite{genel0019} for the topological phase diagram of the system under both periodical and open boundary conditions.
\begin{figure}[t]
\includegraphics[width=4.5cm]{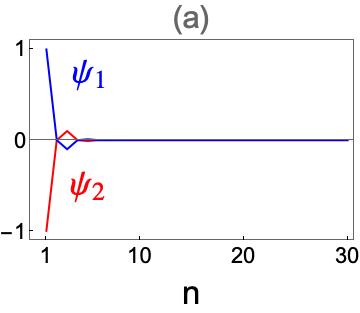}
\includegraphics[width=4.5cm]{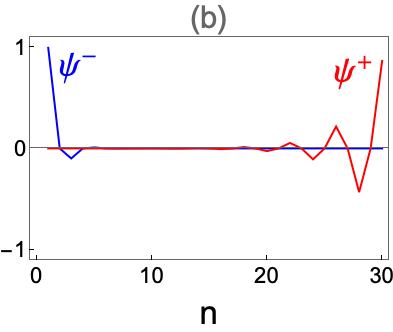}
\caption{The wave packets of the topological zero energy edge states for the non-reciprocal lattice with the parameters $t_1=0.5$, $t_1^{\prime}=0.1$, $t_2=1$ and $N=30$. $\ds{\psi_{1}}$ (blue) and $\ds{\psi_{2}}$ (red) are topological zero energy states localized around the left edge due to the non-Hermitian skin effect. In (b), we plot the superposition states $  \ds{   \psi^{\mp}=P_0( \psi_{1}\mp   \psi_{2} )     }  $, where $P_0$ is a constant satisfying $\sum_n  | \psi^{\mp}|^2=1$ ($P_0=1/2$ for $\ds{\psi^-}$ while $P_0=7.6 \times10^4$ for $\ds{\psi^+}$).  }
\label{thflks0938}
\end{figure}
\begin{figure}[t]
\includegraphics[width=4.2cm]{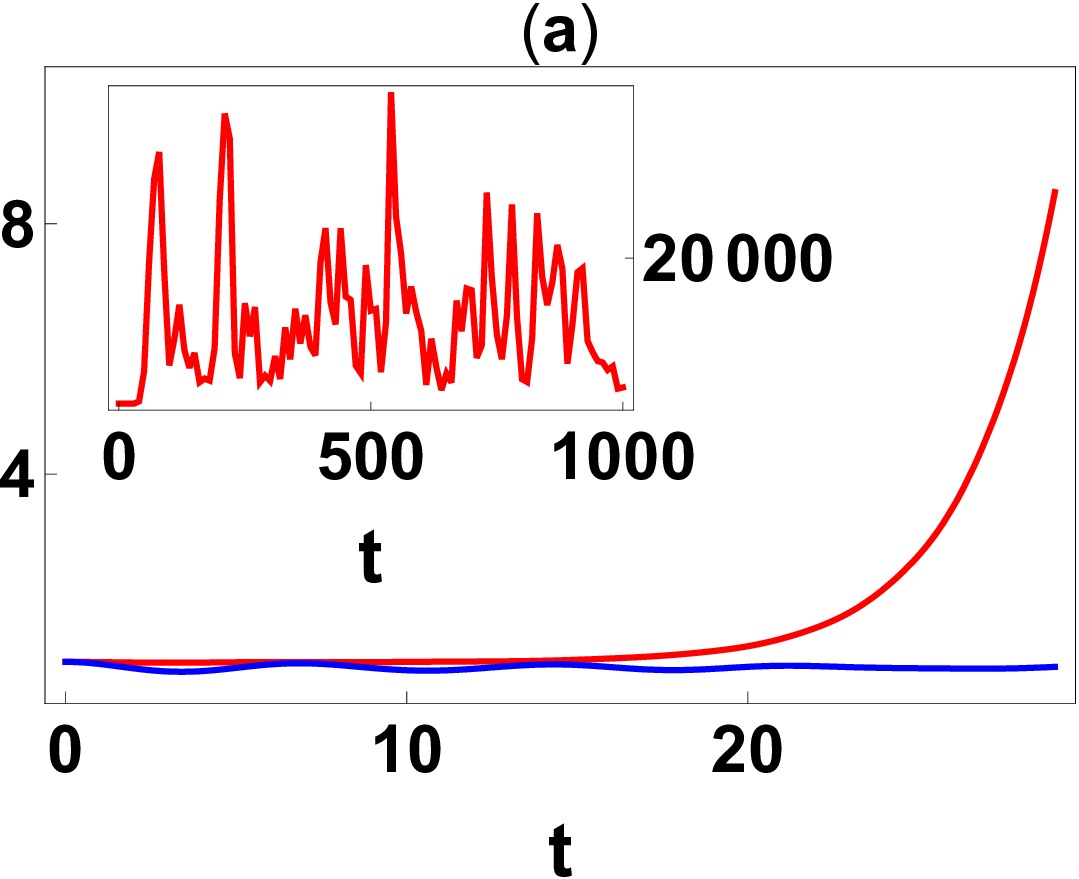}
\includegraphics[width=4.25cm]{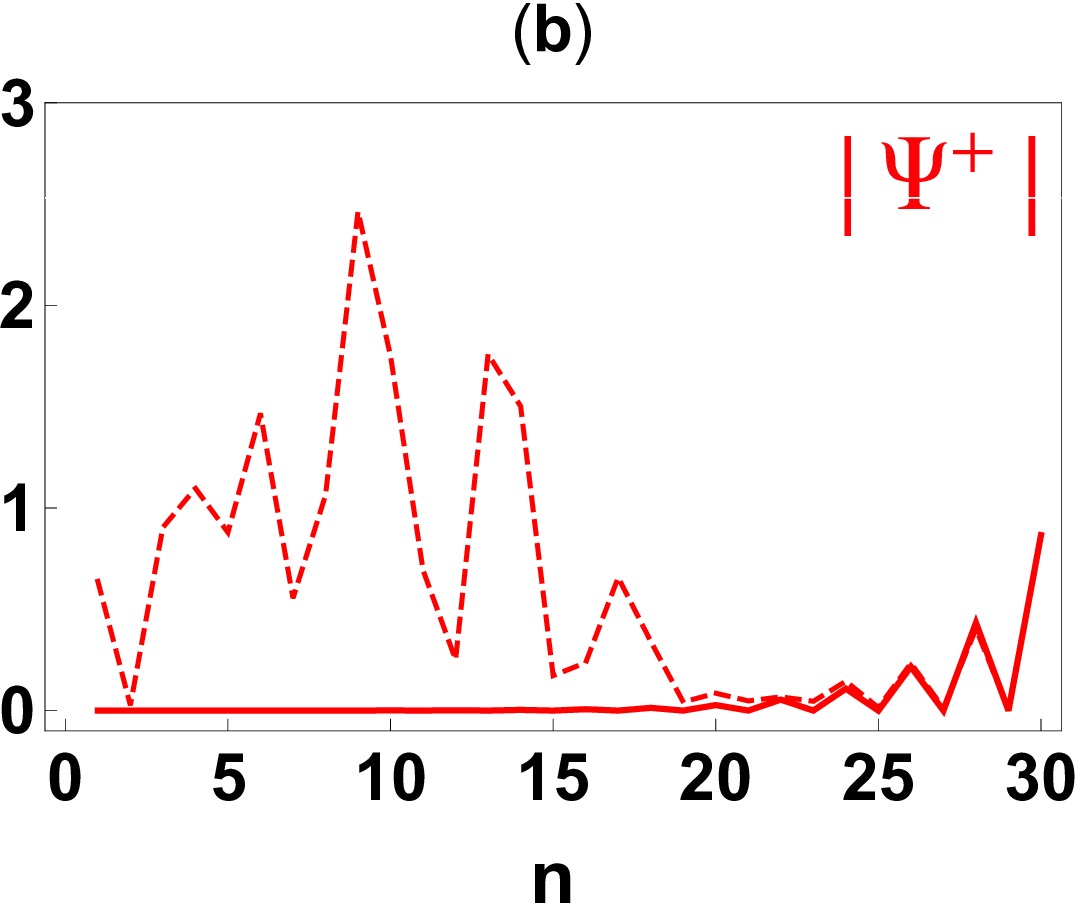}
\caption{The total density $\ds{P_{\mp}(t)=\sum_{n} |\Psi^{\mp}(t) |^2}$ as a function of time for $\Psi^+$ (red) and $\Psi-$ (blue) in the presence of the disorder in (a). The density for the symmetric state grows in time at small times. But it does not grow unboundedly. Instead, it makes non-periodical oscillation with large amplitude as can be seen in the inset, which shows $\ds{P_{+}(t)}$ up to quite large times. The density profiles of $\Psi_n^+$ at $t=0$ (solid red) and $t=25$ (dashed red) are given in (b).}
\label{thflks09382d}
\end{figure}\\
To study topological edge states, consider that our system with an even number of lattice sites $\ds{N}$ is subjected to the open boundary conditions. It is well known that in the topologically nontrivial Hermitian case, $\ds{  t_1^{\prime}=t_1<t_2}$, there exists two topological zero energy eigenstates symmetrically localized around both edges. In the topologically nontrivial non-Hermitian case, the zero energy eigenstates can still appear. However, the non-reciprocity of the hopping amplitudes breaks the spatial symmetry, which implies that they are no longer symmetrically localized around the edges. In fact, not only topological eigenstates but also all of the bulk eigenstates move towards the same edge due to the non-Hermitian skin effect. Let $\ds{\psi_{1}}$ and $\ds{\psi_{2}}$ be the topological zero energy eigenstates. In Fig.\ref{thflks0938}(a), we plot them for the parameters $t_1=0.5$, $t_1^{\prime}=0.1$, $t_2=1$ and $N=30$. As can be seen from the figure, they are localized around the same edge. To check their topological robustness, we introduce disorder in hopping amplitudes for which the chiral symmetry of the system remains intact. The new  hopping amplitudes become $\ds{t_1{\rightarrow}~t_1+{\delta}_{1,n}}$, $\ds{t_2{\rightarrow}~t_2+{\delta}_{2,n}}$ and $\ds{t_1^{\prime}{\rightarrow}~t_1^{\prime}+{\delta}_{3,n}}$, where $\delta_{1,n}$, $\delta_{2,n}$ and $\delta_{3,n}$ are site-dependent and real-valued random set of constants in the interval $(-0.1,0.1)$. Therefore, the hopping amplitudes between neighbouring sites become completely independent. In this case, the eigenstates are deformed but their eigenvalues are still equal to zero, which is a direct result of the topological protection. \\
It is commonly believed that no topological state exists at either edge as a result of the non-Hermitian skin effect. This statement is true for topological eigenstates but one can still construct a topological state localized around the edge where none of the eigenstates is localized. To see this in our system, consider the following symmetric and antisymmetric superpositions of the topological eigenstates
\begin{equation}\label{mkxz294d}
 \psi^{\mp}=P_0( \psi_{1}\mp   \psi_{2} )  
\end{equation}
where $P_0$ is a constant and $\ds{\psi_{1}}$ and $\ds{\psi_{2}}$ are the topological zero energy eigenstates. Note that any linear combination of topological eigenstates is also a topological state since the system is linear. \\
We assume that the total density is equal to one: $\ds{\sum_{n=1}^N  | \psi^{\mp}|^2=1}$. We numerically find that $\ds{P_0=0.5}$ for $\ds{\psi^-}$ while $\ds{P_0=7.6 \times 10^4}$ for $\ds{\psi^+}$ for the parameters used in Fig.\ref{thflks0938}(a). Note this very large numerical difference between them. Such a huge difference is unique to non-Hermitian systems and has an interesting consequence as we will see below. Let us first plot the symmetric and antisymmetric states. In Fig.\ref{thflks0938}(b), we plot them as a function of the site number $n$. One can see that the symmetric state $\psi^+$ (in red) is localized not around the left but the right edge. This clearly shows the possibility of constructing a topological zero energy state localized around the edge where no eigenstate is localized due to the non-Hermitian skin effect. However, $P_0$ becomes very large for such a construction.\\
Let us now discuss time evolution of the topological states (\ref{mkxz294d}). Suppose first that there is no disorder in the system. In this case, no transition among the zero energy eigenstates occur in time and hence the density profiles of the symmetric and antisymmetric states do not change in time. It is well known that the topological zero energy eigenstates $\psi_{1,2}$ conserve their zero energy eigenvalue even in the presence of the symmetry preserving weak disorder (they are topologically protected). A question arises. Can topological protection be still observed for $\ds{ \psi^{\mp}}$? At first sight, this question seems odd because they are just linear combinations of the topological zero energy eigenstates. In Hermitian systems, such symmetric and antisymmetric states remain localized around the edges in the presence of the weak disorder and their zero energy values resist to the disorder. One may naively think that this is also true in non-Hermitian systems. But this is not the case and the picture changes drastically in the presence of even very weak disorder, which is inevitable in a real experiment. To see this clearly, let us start with the initial wave packets $\ds{\psi^{\mp}}$ and  find their time evolutions $\Psi^{\mp}(t)$ ($\Psi^{\mp}(0)=\psi^{\mp}$) in the presence of the  hopping amplitude disorder, which preserves the chiral symmetry. Let us define the time dependent total densities $\ds{P_{\mp}(t)=\sum_{n=1}^N |\Psi^{\mp} (t)|^2}$, where $P_{\mp}(t=0)=1$. In Fig.\ref{thflks09382d}(a), we plot the total densities $P_{\mp}(t)$ as a function of time. As can be seen, the total density oscillates slightly in time for the antisymmetric state (in blue) while it grows in time for the symmetric state (in red). In a non-Hermitian system with real spectrum, a linear combination of two eigenstates with different eigenvalues shows power oscillation (the total power oscillates in time) \cite{powosc}. This is attributed to the nonorthogonality of eigenstates. In our system, $\ds{\psi_{1}}$  and $\ds{\psi_{2}}$ have the same zero energy but some bulk states contribute perturbatively to the total wave packet in the presence of the disorder and consequently we expect slight power oscillation. Therefore, the density oscillation observed for the antisymmetric state  is already expected. However, power growth of the symmetric state is quite unexpected. The Fig.\ref{thflks09382d}(b) shows the density profiles of the symmetric state at two different times. Surprisingly, it is no longer localized around the right edge at large times and becomes extended in time as opposed to the antisymmetric state, which remains localized around the left edge. The same disorder has little effect on one of the superposition state while it has a dramatic change on the other superposition state. This is interesting and has no analog in Hermitian systems. This leads to instability of the topological state $\psi^+$ from the dynamical perspective. This unexpected behaviour can be explained as follows. The constant $P_0$ is of the order of 4 for $\psi^+$ while it is small for $\psi^-$. The presence of the weak disorder makes perturbative changes on the eigenstates $\ds{\psi_{1}}$ and $\ds{\psi_{2}}$. This perturbative change is amplified by 4 orders of magnitude for $\psi^+$. This implies that weak disorder effectively becomes very large disorder only for the symmetric state. In other words, the symmetric topological state is extraordinarily sensitive to the noise in the system. It is well known in Hermitian topological systems that the topological robustness can be observed in the presence of symmetry protecting disorder weak enough not to close the band gap. In a similar way, we think that topological robustness for $\psi^+$ is hard to see in a real experiment because weak disorder effectively becomes very strong. In other words, $\psi^+$ is a topological state with practically missing topological robustness. We stress that the main physics is the same for the symmetric and antisymmetric states. For example, the power oscillation occurs for both of them. However, the large value of the constant $P_0$ makes the oscillation amplitude very large. In the inset of the Fig.2(a), we plot the total density up to a large time. One can see power oscillation with a large amplitude. In a typical experiment, such a large time is not practically realizable and hence one can observe power growth at small experimental times.\\
The first main finding of this paper is that topological robustness can be practically lost for a superpositional topological state in non-Hermitian systems. Below, we study another interesting anomalous feature in a 2D non-Hermitian topological systems.
 
\section{Broken Chirality}
\begin{figure}[t]
\includegraphics[width=5.6cm]{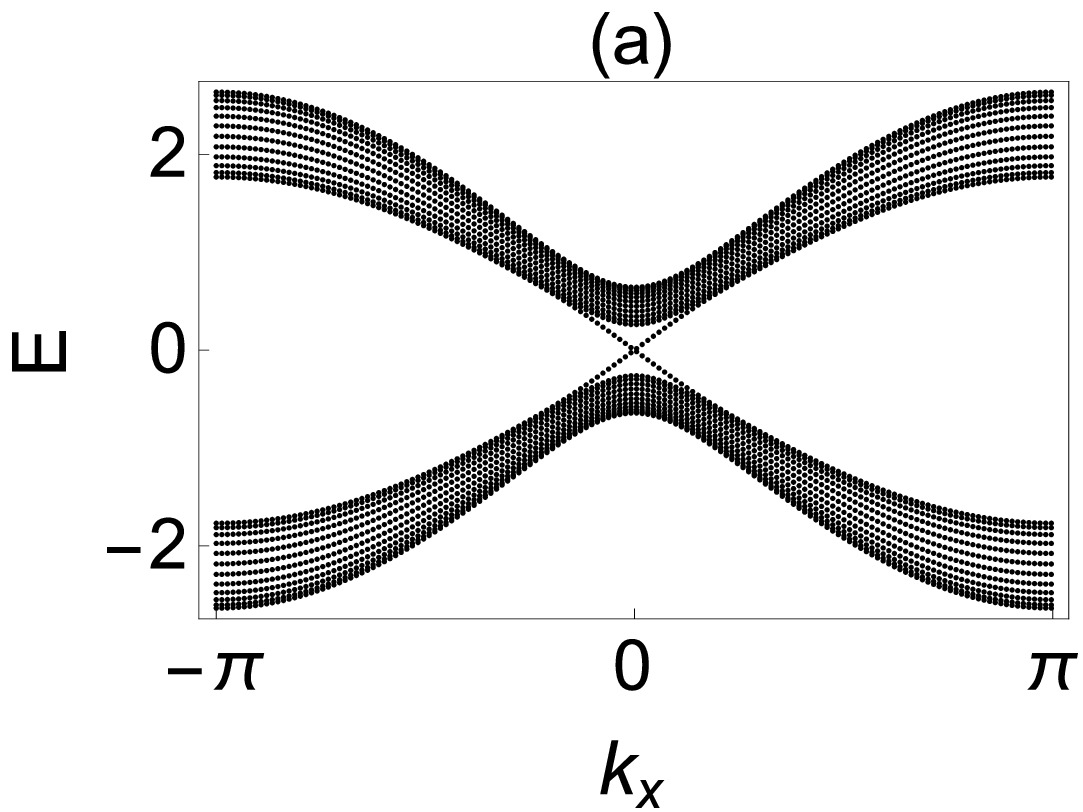}
\includegraphics[width=5.6cm]{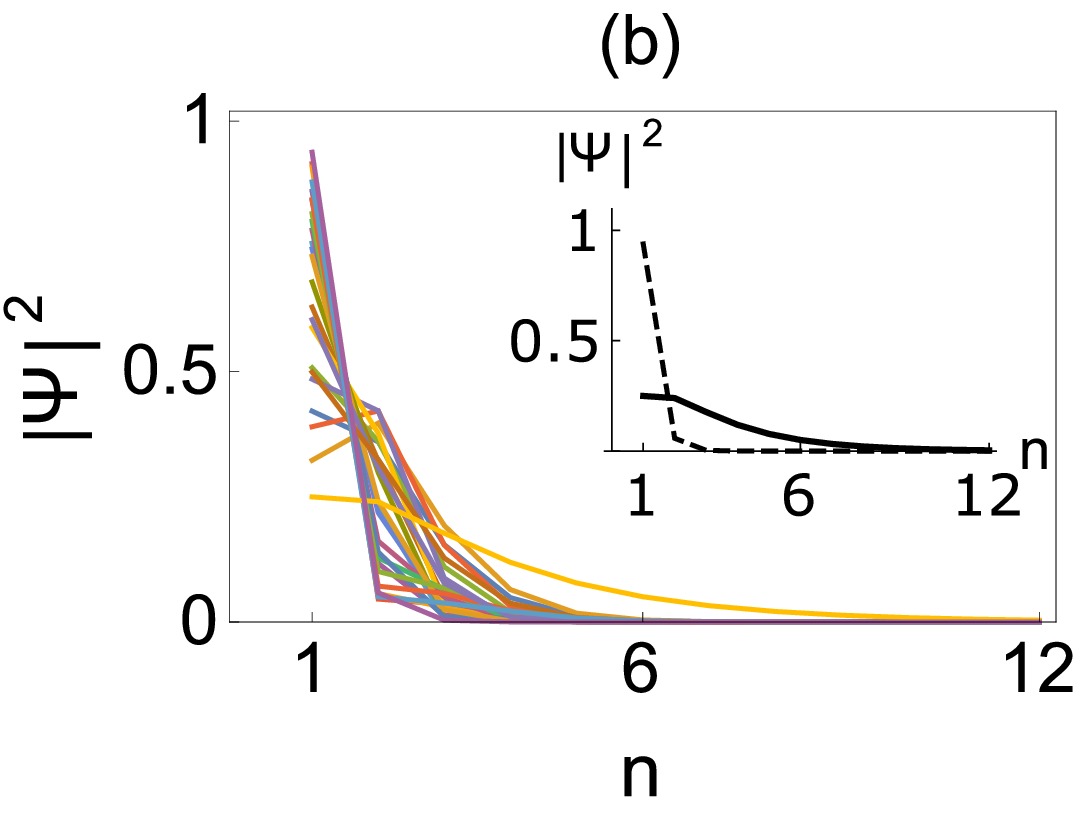}
\includegraphics[width=5.6cm]{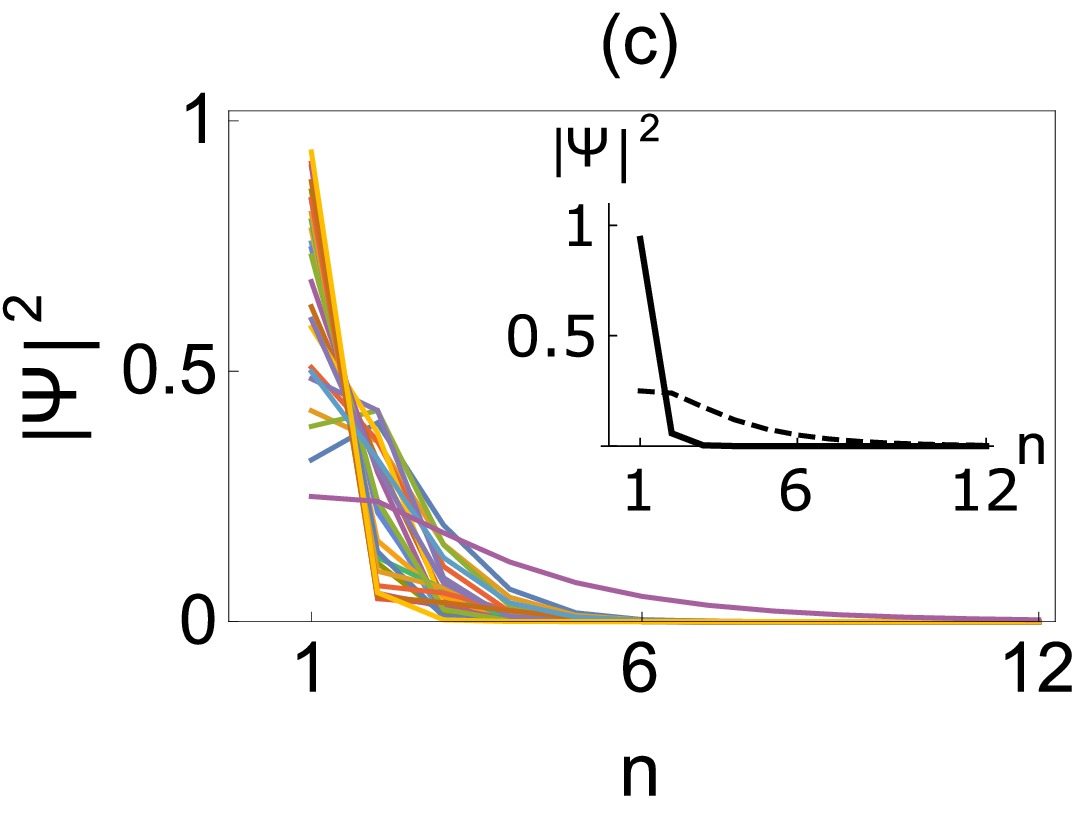}
\caption{The energy spectrum at $\Delta=-1.2$, $A=1$ and  $\alpha=\beta=0.2$ (a). The density profiles for all eigenstates at $k_x=\pi/10$ (b) and $k_x=-\pi/10$ (c). The insets show the density profiles only for the topological states, where the thick and dashed curves are for the states with negative and positive energy eigenvalues, respectively. }
\end{figure}
\begin{figure}[t]
\includegraphics[width=6cm]{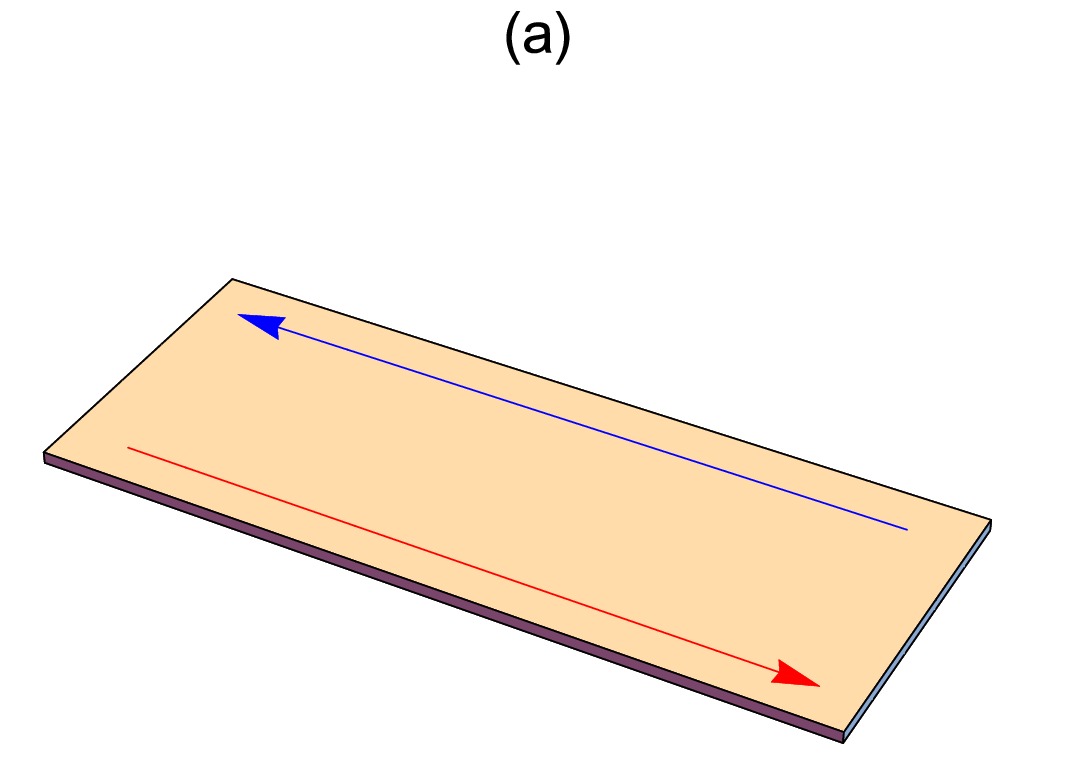}
\includegraphics[width=6cm]{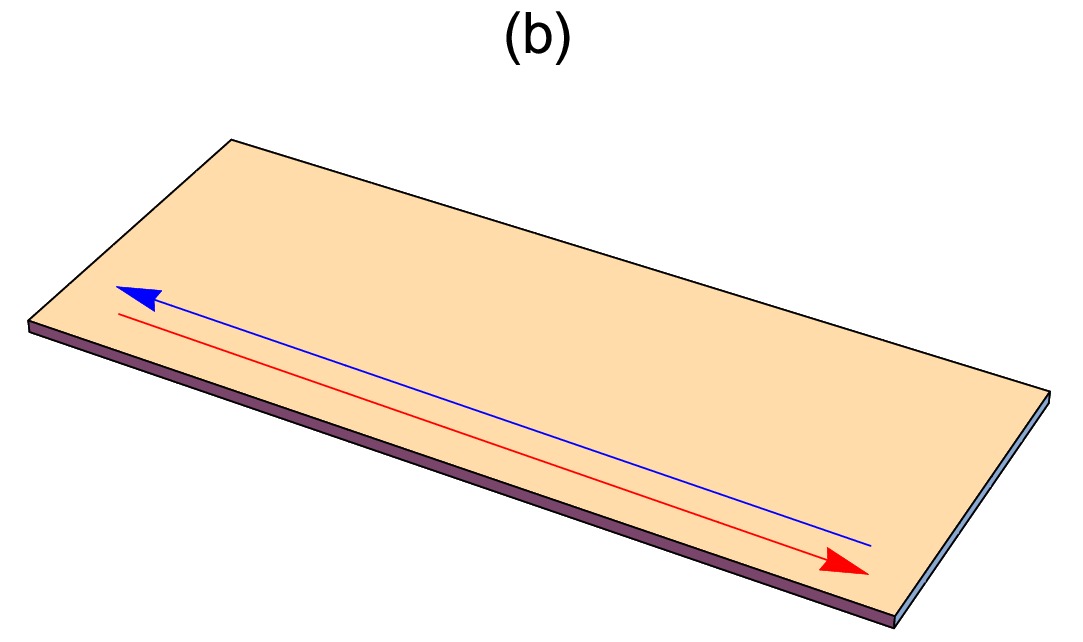}
\label{s6saowa7c}
\caption{Topological states on a 2D strip. In an Hermitian system (a), propagation in one direction only is supported since no state is available at the same energy that propagates in the opposite direction on the same edge. This chirality is the main reason for the suppression of backscattering from defects. However, in the non-Hermitian system, topological states are no longer chiral since forward and backward moving states at the same energy are allowed on the same edge due to the non-Hermitian skin effect (b). Therefore backscattering from defects is possible. In other words, topological states are no longer robust.}
\end{figure} 
\begin{figure}[t]
\includegraphics[width=5.6cm]{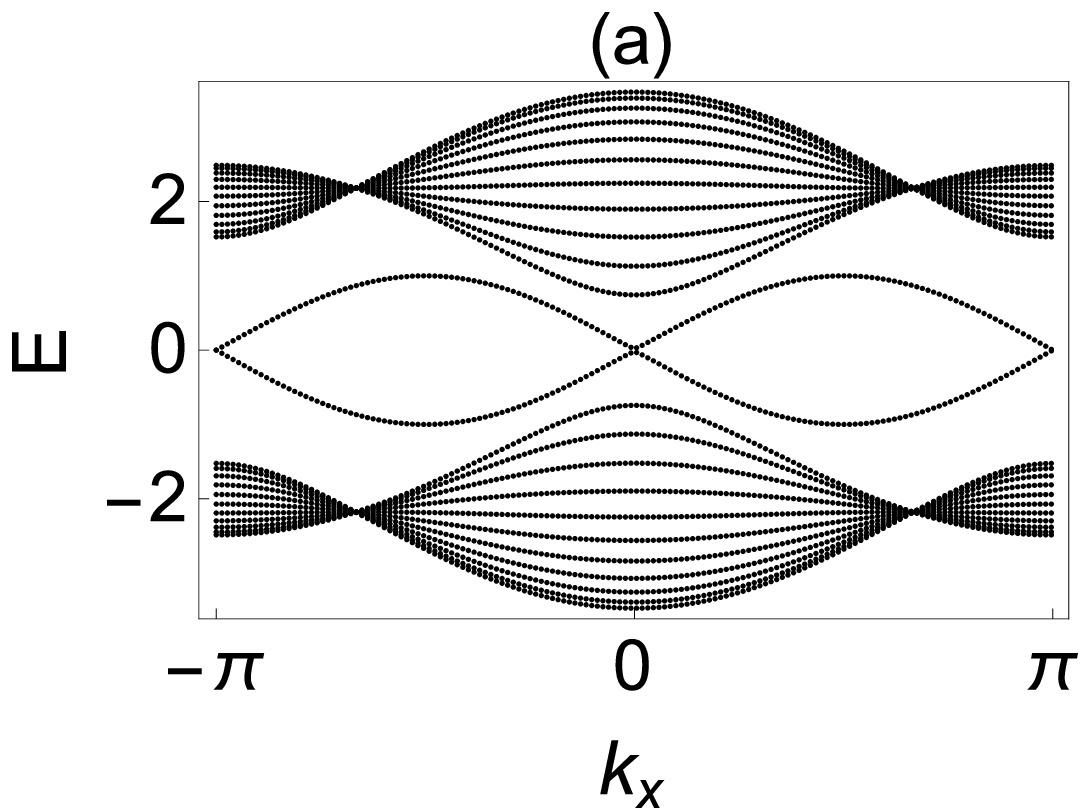}
\includegraphics[width=4cm]{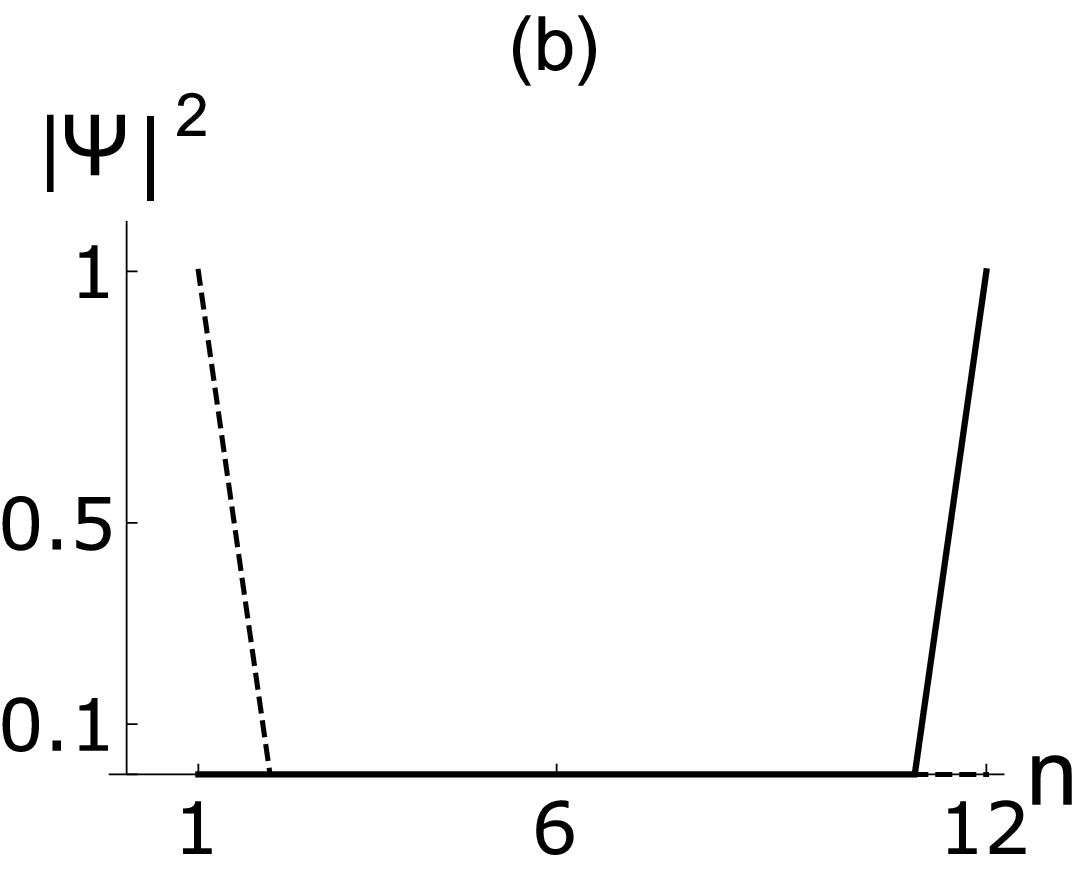}
\includegraphics[width=4cm]{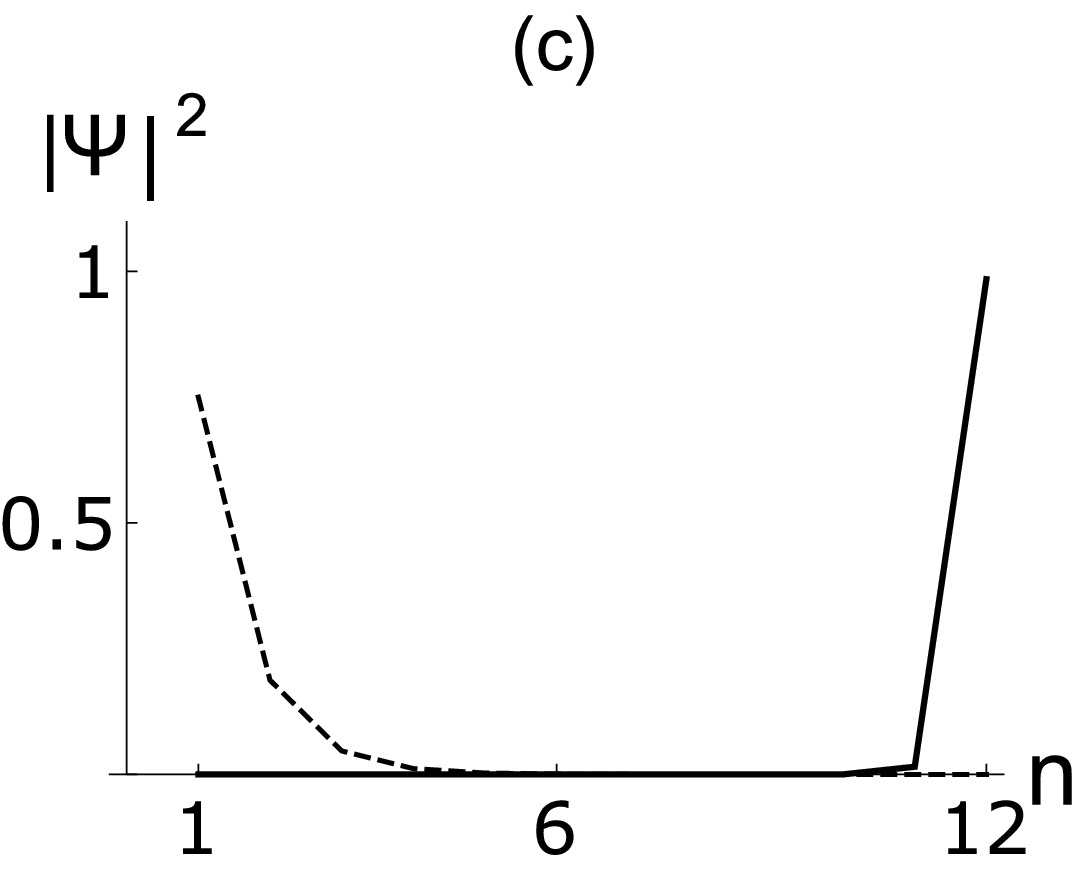}
\includegraphics[width=4cm]{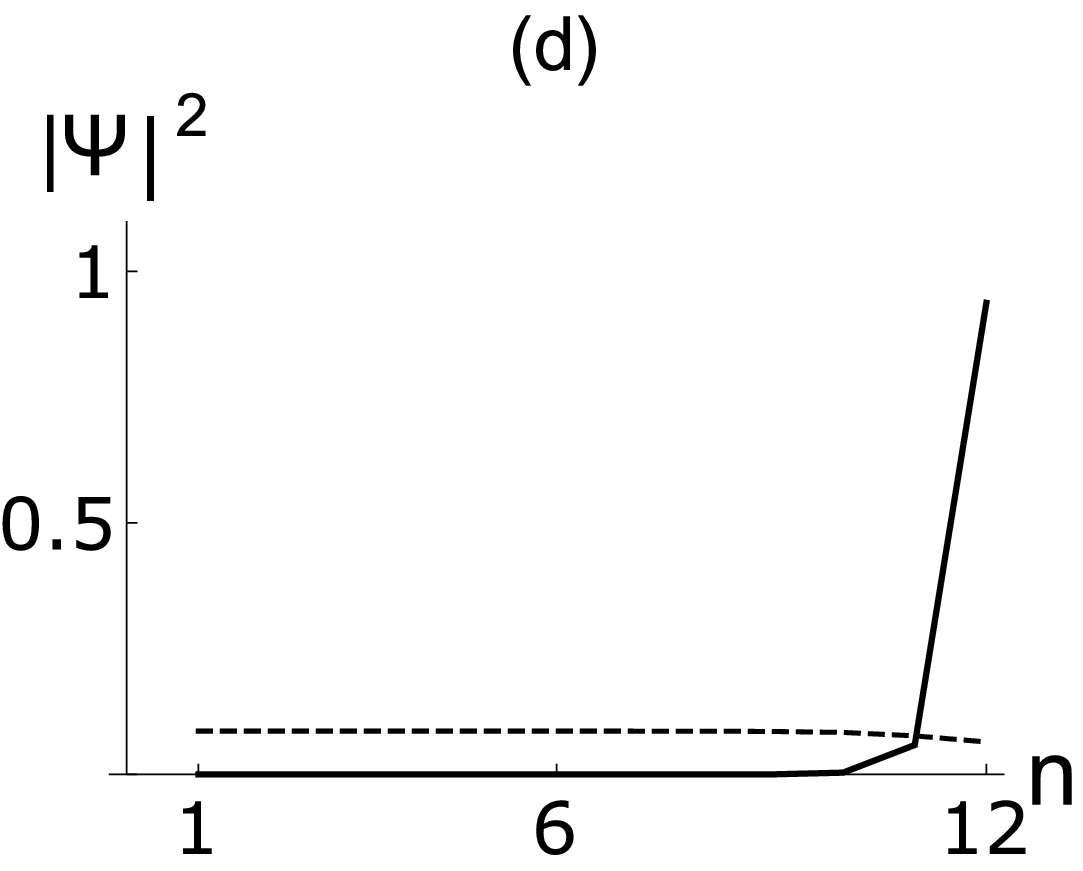}
\includegraphics[width=4cm]{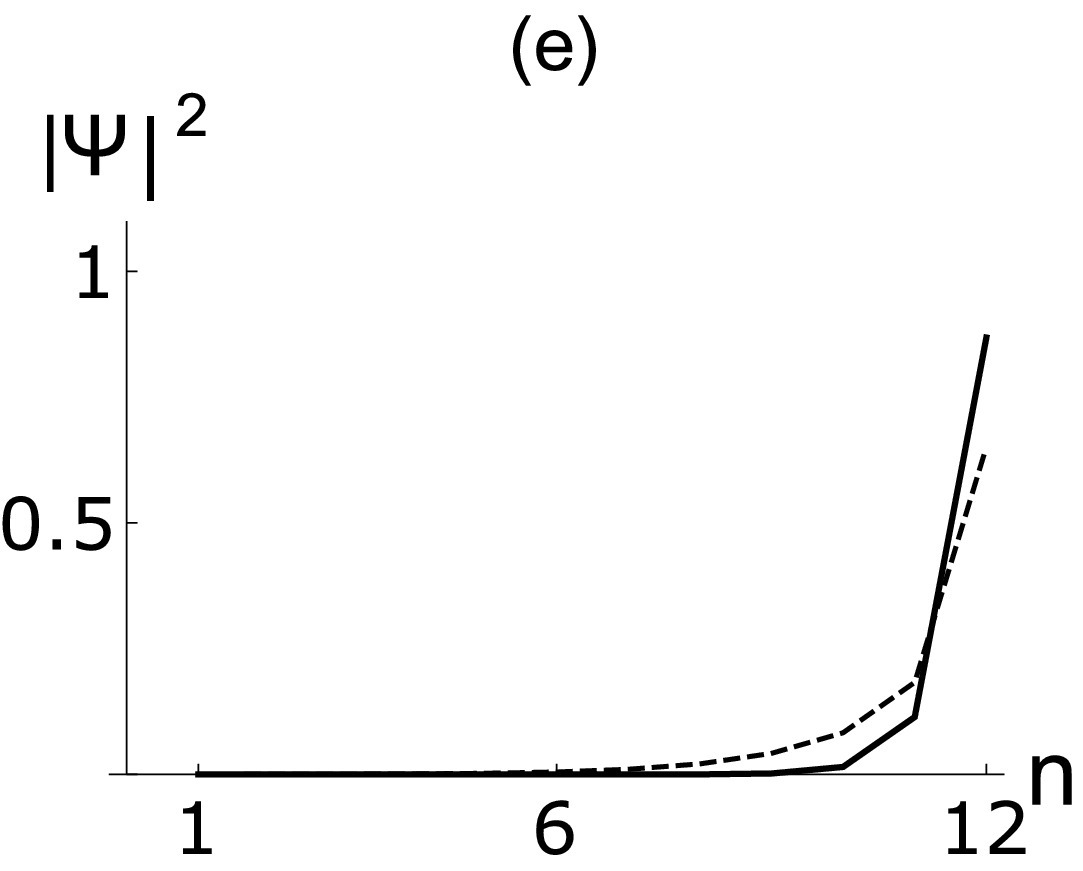}
\caption{The energy spectrum at $\Delta=0.5$, $A=1$ and $\alpha=\beta=4$ (a). The density profiles of the topological edge states at $\ds{k_x=\mp2\pi/3,\mp\pi/2,\mp\pi/3,\mp\pi/10}$ with positive energy eigenvalues for (b,c,d,e), respectively, where the dashed (thick) correspond to positive (negative) values of $k_x$. The localization character of one of the topological edge states is energy dependent while the other one is always localized around one edge. Therefore, chirality of the topological states are also energy dependent.}
\end{figure}
Consider the following non-Hermitian Chern insulator in two dimensions
\begin{equation}
\mathcal{H}  (\textbf{k})= \left( \begin{array}{cc}    \Delta +\cos k_x +F_2   & A   (\sin k_x  -F_1 ) \\ 
A^{\prime}   (\sin k_x  +F_1 )   & - ( \Delta +\cos k_x+F_2 ) 
\end{array}\right) \label{yudj2}
\end{equation}
where $\ds{F_1=\frac{e^{i k_y}-\beta~ e^{-i k_y}}{2}}$, $\ds{F_2=\frac{e^{i k_y}+\alpha ~e^{-i k_y}}{2}}$ and $A$, $A^{\prime}$,  $\alpha$, $\beta$ and $\Delta $ are all real-valued parameters. Noe that this Hamiltonian reduces to the well-known Hermitian Chern insulator Hamiltonian when $A=A^{\prime}$ and $\alpha=\beta=1$ \cite{qwz}. The corresponding energy eigenvalues come in pair and are given by $\ds{E=\mp\sqrt {  ( \Delta +\cos k_x+F_2 )^2-A A^{\prime}   (\sin^2 k_x  -F_1^2 )        }}$. \\
Let us consider a finite system with open edges and study topological states. Suppose that the system is periodical along $x$-direction and has open edges along $y$-direction, which consists of $N$ sites. Since the system is translationally invariant along $x$, we can partially do Fourier transformation and the resulting Hamiltonian has discrete form indexed by the continuous parameter $k_x$. For our numerical computation, we take $A=A^{\prime}=1$, $\Delta=-1.2$, $\alpha=\beta=0.2$ and $N=24$. We find that the corresponding system has fully real spectrum. The Fig.3 (a) plots the energy spectrum as a function of $\ds{k_x}$. One can see the states in the band gap connecting the lower and upper bands across the bulk gap. They are topological states departed from nearly $k_x=\mp\pi/6$. They cross each other at $k_x = 0$ and switch places and enter the upper bulk band at the symmetrical point from where they departed. These states are edge states propagating in opposite directions (at a given $E$). So far, there is nothing interesting since a similar structure can also be seen in the Hermitian counterpart. Let us now plot the density profiles of the edge states at a given $k_x$. In the Fig. 3 (b,c), one can see that the edge states are localized around the same edge as opposed to the Hermitian system, where edge states are localized symmetrically at the opposite edges. In other words, forward and backward moving edge states (with positive and negative $k_x$ values) are localized on the same edge. This implies that the chirality is lost in our system (Fig. 4). What is special in our system is that the edge states are no longer robust against backscattering. In the Hermitian counterpart, the forward and backward moving edge states at the same energy are localized around the opposite edges. Therefore there is no available state moving in the opposite direction on the same edge. The topological states moving oppositely at the same energy are well separated so no transition occurs. This chirality of the edge states is mainly responsible for the robustness of them against backscattering in the presence of the weak disorder in Hermitian systems. But this is not the case in our system as the topological states are localized on the same edge and weak symmetry protecting disorder induces transition between forward and backward moving edge states. Consequently, the topological states have no immunity to backscattering. This is the second main finding of this paper. This poses a question of advantages of topological states in such a system. \\ 
To make further exploration, we plot the energy spectrum when $A=A^{\prime}=1$, $\Delta=0.5$ and $\alpha=\beta=4$ in the Fig. 5.(a). We find that the corresponding spectrum is purely real. Apparently, topological states appear in the band gap for all values of $\ds{k_x}$. In this case, we can scan $k_x$ to study chirality of topological states. In Fig. 5.(b-e), we plot the density profiles for topological states for various values of $\ds{{\mp}k_x}$ with $E>0$. The topological edge state with negative $k_x$ (thick curve) is always localized around the right edge. However, the localization character of the topological state with positive $k_x$ (dashed curve) depends on $\ds{k_x}$. For example, the topological state is localized around the left (right) edge at $\ds{k_x=2\pi/3}$ ($\ds{k_x=-2\pi/3}$) as can be seen from (b). Therefore the topological edge states are chiral at this particular value of $k_x$. As $\ds{k_x}$ decreases, the topological states with positive $k_x$ are shifted towards the right edge as can be seen from (c-e).  At around $\ds{k_x=\pi/3}$ (d), the topological state becomes extended all over the system (recall that it is also delocalized in $x$ direction due to the translational invariance along $x$ ) \cite{anaskef}. For small values of $\ds{|k_x|}$, both states are localized around the same edge as can be seen from (e) and chirality of them are broken. These show that chirality breaking can be energy dependent through the parameter $k_x$. 

\section{Conclusion}

In this paper, we have found two anomalous features of topological states in a non-Hermitian nonreciprocal system. Firstly, we have discussed that a superposition of topological states can be practically fragile against the disorder in a non-Hermitian system even if the topological eigenstates are robust. This superposition-induced loss of topological protection is unique to non-Hermitian systems. We further find power growth effect at small times for a superposed eigenstates in a non-Hermitian system with real spectrum. Secondly, we have shown that chirality of topological edge states can be lost in two dimensional non-Hermitian systems as a result of the non-Hermitian skin effect. This is also unique to non-Hermitian systems and can lead to the loss of robustness of topological edge states. We have discussed that chirality of topological edge states in a non-Hermitian two dimensional system can be energy dependent. We think that our findings will pave the way for the understanding of topological phase in non-Hermitian systems.

\end{document}